\documentclass{article}  
\usepackage{lineno,hyperref}

 \usepackage{setspace}
\usepackage{tocloft}

\usepackage{amsmath,amssymb,amsfonts}
\usepackage{color}
\usepackage{enumerate}
\usepackage{graphicx}   
\usepackage{epstopdf}
\DeclareGraphicsExtensions{.pdf,.eps,.jpg,.mps}
\usepackage{subcaption}
 
\usepackage{hyperref}
\hypersetup{
 colorlinks,
 citecolor=blue,
 }
\usepackage{url}

\def\red{\textcolor{black}}

\title{Visualization of latent and depleted fingermarks on CDs and DVDs using columnar thin films}

\author{Muhammad Faryad$^{a,b,\ast}$, Akhlesh Lakhtakia$^a$, \\Ricardo A. Fiallo$^a$, and Partha P. Banerjee$^c$\\\\
$^a$The Pennsylvania State University, \\Department of Engineering Science and Mechanics, \\University Park, PA 16802-6812,   USA.\\
$^b$Lahore University of Management Sciences, \\Department of Physics, Lahore 54792, Pakistan.\\
$^c$University of Dayton, Department of Electro-Optics and Photonics, \\Dayton, OH 45469, USA.\\
$^\ast$\texttt{muhammad.faryad@lums.edu.pk}}

\begin{document}

\maketitle

\begin{abstract}
CDs and DVDs are forensically  important substrates for latent fingermarks. Upright columnar thin films (CTFs) grown conformally over  these substrates were shown to significantly enhance the visual quality of fingermarks. Enhancement to  maximum possible grade for visual quality occurred, even for the samples that were developed three days after the fingermarks were placed on them. The CTFs were deposited by directing a collimated vapor flux of an evaporant material towards the substrate placed at an angle and made to rotate rapidly. 
 \end{abstract}




\section{Introduction}
By the twenty-fifth birthday of compact discs (CDs) in 2007, more than 200 billion CDs and digital video discs (DVDs) had been sold in the world \cite{BBC2007}. Even though the sales of CDs and DVDs have declined with the rise of streaming services since then \cite{Statista}, they continue to have a ubiquitous presence in homes and workplaces, which makes them  forensically  important substrates for  latent fingermarks.  Latent fingermarks are difficult to visualize with the naked eye. Several techniques have been devised over the last century to enhance  latent fingermarks for visualization by fingermark examiners. The most common fingermark-development techniques include dusting with a powder \cite{Lee}, chemical reactions with appropriate reagents \cite{Lennard}, immersion in a small particle reagent (SPR) \cite{Haque},  vacuum metal deposition \cite{Batey}, and cyanoacrylate fuming \cite{Watkin}. Imaging with infrared \cite{Crane} or X-rays \cite{Worley} exploits  a specific chemical species in the  fingermark residue. 

Powder dusting,   cyanoacrylate fuming, and the SPR technique have  been used by Jasuja \textit{et al}  \cite{Jasuja} to develop latent fingermarks on CDs. However, their focus  was on preservation of electronic data after fingermark development. If the goal is to preserve the fingermark for a long time and not the electronic data on CDs, then the columnar-thin-film (CTF) technique appears desirable for visualization of latent fingermarks \cite{Shaler2011,Lakhtakia2011}. In this technique,  an upright CTF of an appropriate material is deposited on top of a rapidly rotating substrate with the fingermark, this development process taking place in a vacuum chamber   \cite{Steve2014}, the apparatus being the same as   used for  vacuum metal deposition \cite{Steve2015}.  

\red{CDs and DVDs are  novel substrates for the CTF technique. This fingermark-visualization technique has been tried with varying degrees of success on sheets of acrylonitrile butadiene styrene, Nylon 66 (which represents polyamides), and soft polymers used to make bags and tapes \cite{Muhlberger2014}. But CDs and DVDs are made of layers of clear polycarbonates, which have different chemical composition and physical characteristics.} 

\red{Many types of CDs and DVDs have at least one internal interface which has spiral tracks of nanoscale bumps and is coated with a metallic thin film (to reflect a laser beam) \cite{Canada}. These features  endow the CD/DVD with the optical reflection characteristics of a metallic grating \cite{Loewen}, which makes it  difficult for a forensic examiner to clearly see fingermarks on the exterior surface. Writable and rewritable CDs and DVDs also contain a dye (usually, an azo dye, a cyanine dye, or phthalocyanine) which reflects  light only in the blue-green portion of the visible spectrum \cite{Canada}, and this characteristic can also make it difficult to clearly see fingermarks. Parenthetically, CDs and DVDs are two different types of substrates because DVDs have denser arrays of smaller bumps.}  

Therefore, we set out to investigate the suitability of the CTF technique for the development of latent and depleted fingermarks on commonly used writable CDs and DVDs (CD-R,  DVD-R) and DVD-ROMs, which are usually used for the distribution of  movies and software.
A latent fingermark  is   left on a substrate, such as the exterior polycarbonate surface of a CD or DVD, after the finger has been rubbed on one's face and/or waiting long enough for the finger to become coated with sebum. A depleted fingermark  is defined to be the one that is  left immediately right after  the latent fingermark, without waiting or rubbing the finger anywhere again. The   fingermark-acquisition method, sample aging, photography, grading scheme,  CTF deposition, and imaging with a scanning electron microscope (SEM) are explained  in Sec. \ref{methods}. The results are presented and discussed in Sec. \ref{results}.

\section{Materials and Methods}\label{methods}
\red{We used a singly writable CD (CD-R, Memorex\textsuperscript{\textregistered}, $700$ MB, $52$X Multispeed), a singly writable DVD (DVD-R, Verbatim$^{\rm TM}$, $4.7$ GM, $16$X Speed), and a read-only DVD (DVD-ROM of the movie \textsc{Free Birds}, 2013) for all the results presented here.
}
A set of 24 fingermarks was obtained from  a specific finger of a single donor. Both CDs and DVDs were cut into smaller pieces using a shear machine and were cleaned with an alcohol wipe before acquiring the fingermarks. Every latent fingermark was acquired after the donor rubbed the chosen finger on their face and behind the ears for 10 s. A depleted fingermark was obtained right after the latent fingermark, without waiting or rubbing the finger anywhere.
Both the latent and the depleted fingermarks were aged for either one day or three days in the typical  office atmosphere (temperature $\sim$ 21~$^\circ$C, relative humidity $\sim$ 20-40\%) before CTF deposition.  

CTFs of either (i) chalcogenide glass of nominal composition Ge$_{28}$Sb$_{12}$Se$_{60}$ or (ii) nickel were deposited in a vacuum chamber with a pressure of $85\pm15$~$\mu{\rm Torr}$. Chalcogenide glass was ground to a fine powder and nickel was used in the form of a wire. The deposition parameters  given in Table \ref{table1} were guided by previous research \cite{Nagachar2020}.  Either  chalcogenide glass or nickel was resistively heated by passing an electric current through an appropriate boat \red{containing that material}
and the collimated vapor flux thereby generated was directed towards the substrate that was tilted at an angle of  $20$~deg with respect to the arriving vapor flux.  The substrate was rotated at a rate of $180$~rpm around an axis passing through the center of the substrate holder, while the CTF deposition rate was maintained in a narrow range. The sample was left to cool in the chamber for about an hour after the deposition before venting the chamber. 

\red{The substrate tilt angle of $20$ deg and the rotation rate of $180$ rpm were chosen based on  earlier optimization studies on 23 different forensically relevant substrates \cite{Muhlberger2014}. CTFs of several different thicknesses ranging from 50 nm to 1500 nm were deposited.} The optimal value of the CTF thickness \red{for visualization}  was found to be  $1000$~nm for chalcogenide glass and $100$~nm for nickel. 

\begin{table}[h]
 \caption{Materials, boats, and deposition parameters. }
    \centering
    \begin{tabular}{ccccccc}
	\hline
	 Evaporant Material  & Boat & Current   &  Rate  & CTF Thickness \\
	    &   &   (A) &    (nm/s) &   (nm)\\
	\hline
	 Chalcogenide glass  & S22-.005W & 65 & 0.5-1.5& 1,000\\
	Nickel & S4-.005W & 75 & 0.05-0.1  & 100\\
	\hline
    \end{tabular}
    \label{table1}
\end{table}

Immediately before and immediately after CTF deposition, every sample was imaged by a Nikon D90 camera with a Nikon  105 mm lens. The mode dial was  set to close-up pictures. Most of the pictures were taken with autofocus except when the sample was too shiny for the auto-focus mode to work.  
The sample was illuminated with a Sylvania 100 W incandescent bulb at an angle that gave the best possible image of the fingermark for each sample. The sample-to-camera distance   was kept the same for all samples.  

\red{The photographs of latent and depleted fingermarks before CTF deposition invariably appear to be slightly out of focus due to the unavoidable diffraction of light from the interior metallic grating. In Sec.~\ref{results}, we  present the best images of fingermarks before CTF deposition for comparison with the post-deposition images, which were easy to obtain since the effect of grating diffraction was either (i) eliminated by reflection at the air/CTF interface for  nickel CTFs or (ii) de-emphasized by reflection by the air/CTF and CTF/residue interfaces for chalcogenide-glass CTFs.}

Every one of the 24 fingermarks was graded \red{by one of us (M.F.)} for visual quality before and after CTF deposition  with a grading scheme used extensively by the UK Home Office \cite{Sears}. According to this scheme, the fingermark is given an integer grade from $0$ to $4$. A grade of $0$ indicates no evidence of a fingermark whereas a grade of $4$ indicates  complete ridge detail of the whole fingermark. A grade of $1$ indicates  evidence of contact of the substrate with the finger but no visual evidence of the ridges. A grade of $2$ indicates that ridges are visible in about a third of the fingermark, whereas a grade of $3$ indicates the ridge visibility for about two thirds of the fingermark. Only a fingermark with a grade of $3$ or $4$ can be used to reliably identify the person.

The cross-sectional images of the deposited CTFs were obtained with a field-emission SEM (Verios G4 UC, Thermo Fisher Scientific, Millersburg, PA). For this purpose, the sample was flash frozen using liquid nitrogen before being cut with a handheld shear cutter. The freezing process allows the samples to fracture while maintaining the integrity of the CTF structure.  The sample was then mounted   on a standard $90$-deg stub with double-sided carbon tape in such a way that the cleaved edge would lie perpendicular to the lens for imaging. The samples were then sputter coated using an iridium target (Emitech K575X, Fedelco,   Madrid, Spain) for $35$~s at  $40$-mA current. Images were taken in immersion mode using a through-the-lens detector, with $3$-kV voltage, $0.10$-nA current/spot size, and $2.5$-mm working distance, which allowed for high-magnification imaging of the CTFs.

\begin{figure}[htbp]
\centering
\includegraphics[width=1.0\linewidth]{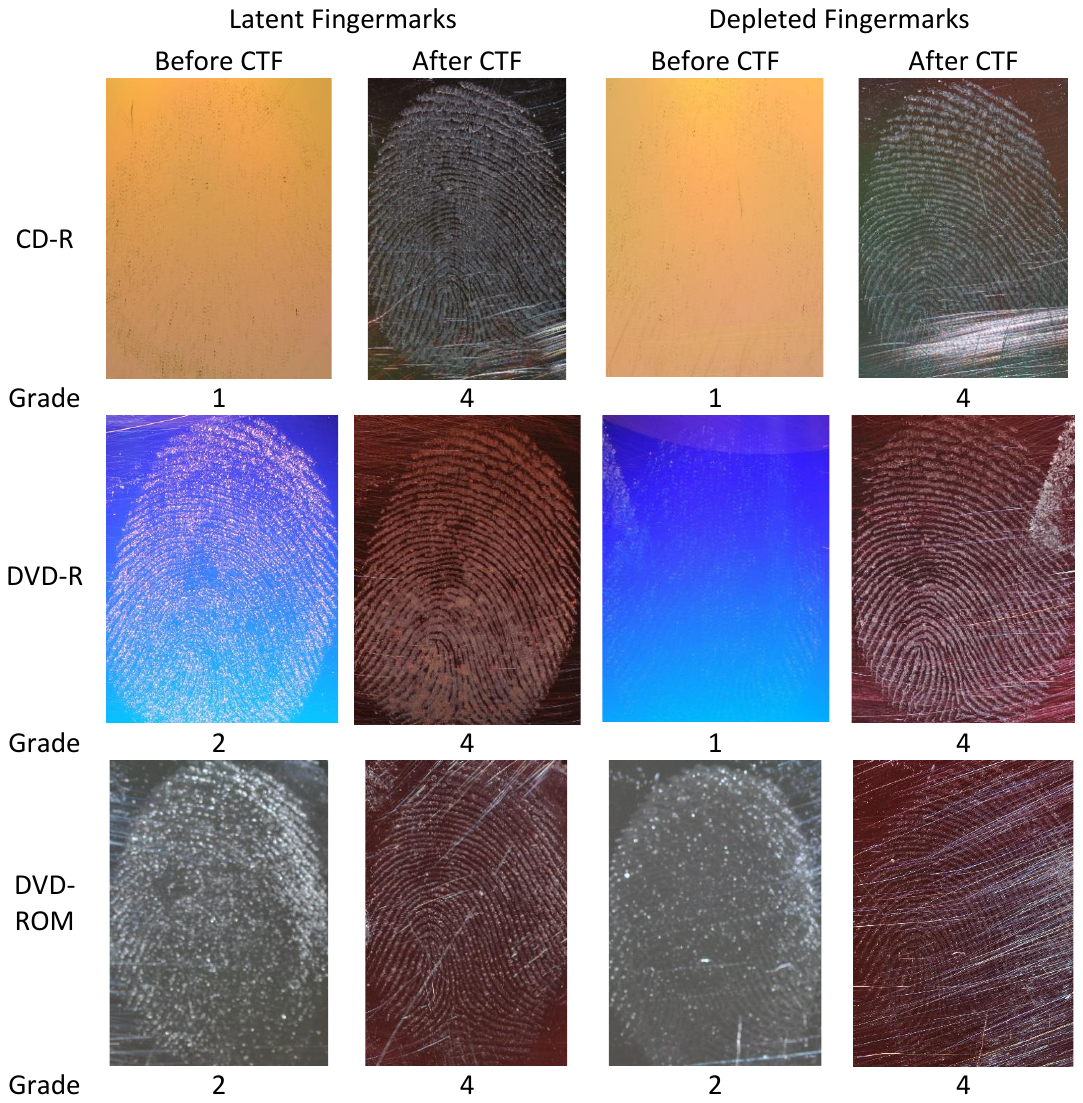}
\caption{ Photographs of latent and depleted fingermarks on CD-Rs, DVD-Rs, and DVD-ROMs before and after the deposition of a $1000$-nm thick CTF of chalcogenide glass.  The fingermarks were aged for $24$~h in the office environment before CTF deposition. The UK Home Office quality grade is provided below each photograph. }
\label{fig1}
\end{figure}

\begin{figure}[htbp]
\centering
\includegraphics[width=1.0\linewidth]{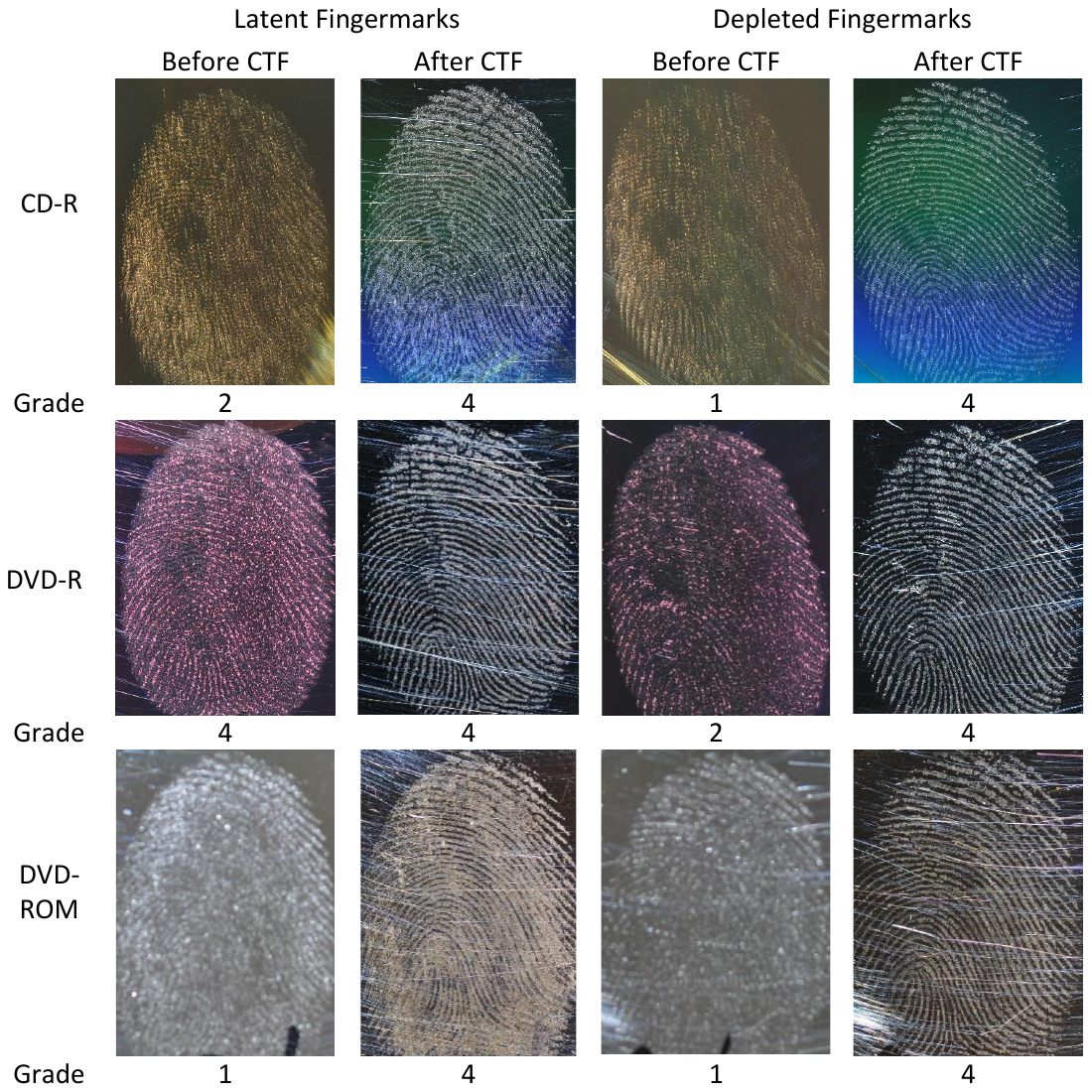}
\caption{Same as Fig. \ref{fig1} except that a nickel CTF of $100$ nm thickness was deposited on each sample.}
\label{fig2}
\end{figure}

\begin{figure}[htbp]
\centering
\includegraphics[width=1.0\linewidth]{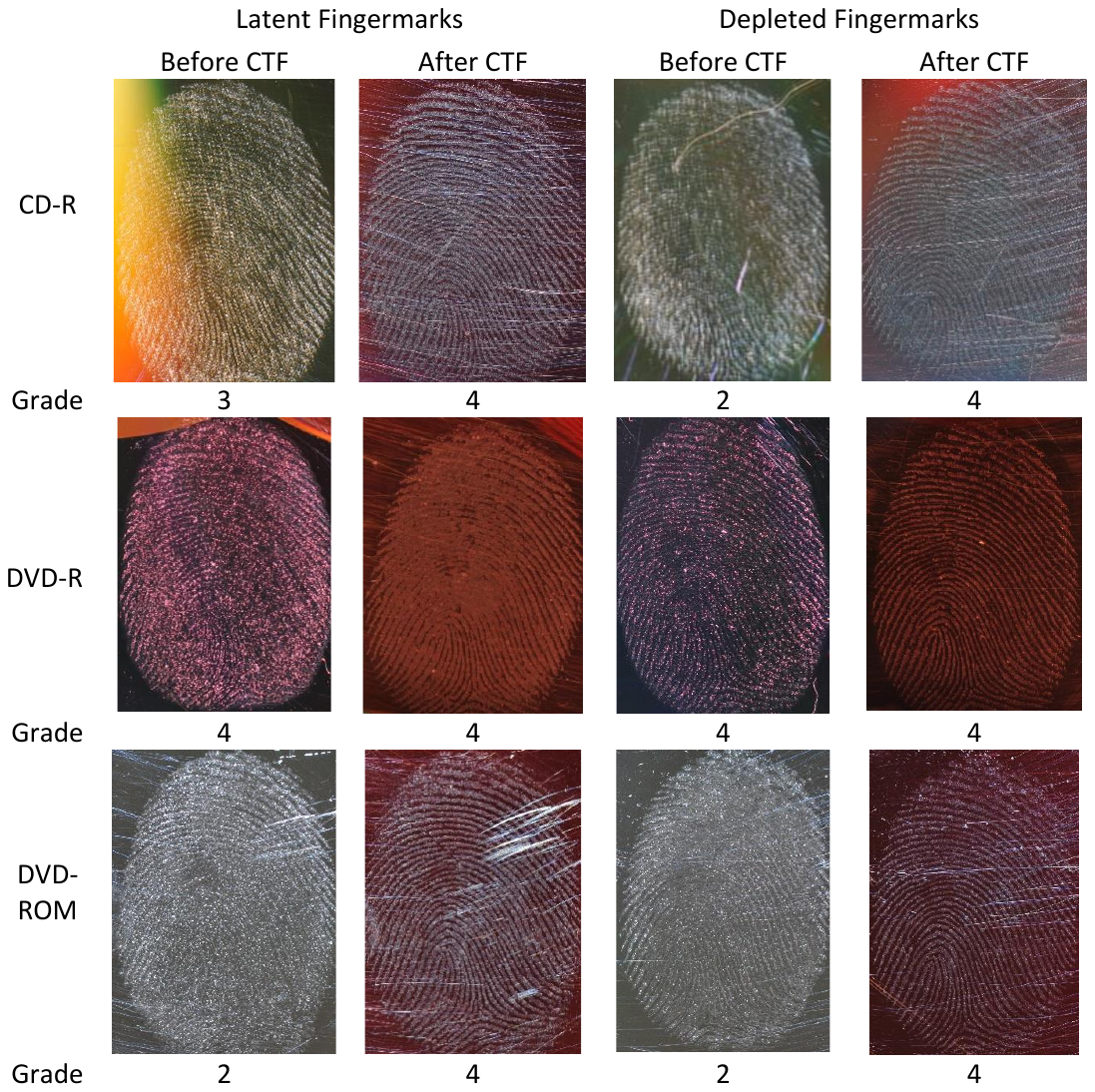}
\caption{Same as Fig. \ref{fig1} except that the fingermarks  were aged for $72$~h    before CTF deposition.}
\label{fig3}
\end{figure}

\begin{figure}[htbp]
\centering
\includegraphics[width=\linewidth]{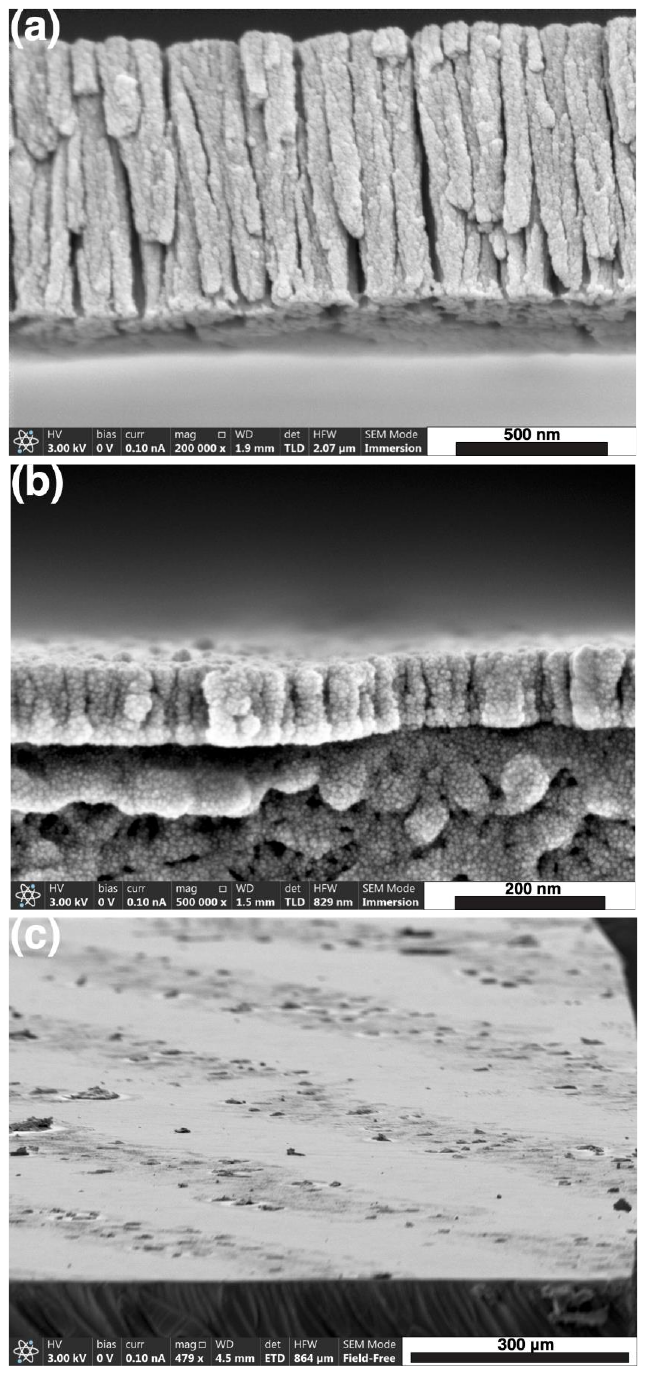}\caption{(a,b) Cross-sectional and (c) tilted top-surface}
scanning-electron micrographs of   CTFs of  \red{(a) chalcogenide glass  and (b,c) nickel deposited on fingermarks on   DVD-Rs.}
\label{fig4}
\end{figure}

\section{Results and Discussion}\label{results}
Photographs of six samples aged for one day before CTF deposition and developed with a $1000$-nm-thick CTF
of chalcogenide glass are presented in Fig. \ref{fig1}.  None of the six fingermarks received a quality grade exceeding $2$ before
CTF deposition. Furthermore, the depleted fingermark on any specific substrate was of either the same or lower visual quality than the latent fingermark on the same substrate, per expectation \cite{Williams2015}. Thus, none of the six fingermarks could be used to reliably identify the person.  However, all fingermarks had a quality grade of $4$ after  development by CTF deposition, this being excellent for the purpose of  identification.

Figure \ref{fig2} presents photographs of six fingermarks aged for one day before CTF deposition and developed with a $100$-nm-thick CTF of nickel. Again, the depleted fingermark on any specific substrate was of either the same or lower quality than the latent fingermark on the same substrate. Also,
 all fingermarks received a quality grade of $4$ after    CTF deposition.

Photographs of six samples aged for three days and developed with chalcogenide glass are presented in Fig. \ref{fig3}. The trends identified for the previous 12 fingermarks are confirmed, as also for the six
fingermarks aged for three days and developed
with a $100$-nm-thick nickel CTF (photographs not shown).
All 12   fingermarks had a grade of $4$ after      CTF deposition. 

SEM images of the deposited CTFs showed the same morphology regardless of the type of the electronic media. Figure \ref{fig4}(a) presents a cross-sectional image of a   chalcogenide-glass CTF deposited on a fingermark on a DVD-R. The morphology is typical of a CTF deposited on a fast-rotating and tilted substrate \cite{Steve2014}. Shadowing  of the collimated vapor flux by  columns is responsible for columnar morphology, whereas rapid substrate rotation during deposition makes the columns grow upright. 
This upright growth of columns  helps replicate the topography of the substrate making the faithful development of the fingermarks possible. 
Figure \ref{fig4}(b) shows a cross-section of a nickel CTF deposited on a fingermark  on a DVD-R. Again, the figure shows similar qualitative features as that of the CTF made by evaporating chalcogenide glass. Therefore, different types of evaporant materials (metal or glass) deliver dense, upright, columnar morphology on fingermarks.  

\red{Figure \ref{fig4}(c) shows a tilted top-surface image of a nickel CTF grown over a fingermark on a DVD-R. Clearly, the underlying ridge-and-valley topography of the  fingermark has been reproduced by the CTF.}
As the local reflectance of a thin film depends on the optical thickness of the fingermark residue, and the ridges are several optical wavelengths in width, a visual contrast is created between the CTF deposited upright on  a ridge and the CTF deposited upright in an adjacent valley. This  contrast is responsible for enhancement of visual quality for identification.
 
 \red{The difference in optimal thickness of nickel and chalcogenide-glass CTFs by an order of magnitude is due to the different electromagnetic mechanisms   responsible for the enhancement of quality grade for visualization. Nickel is a metal and the 100-nm thickness of the nickel CTF exceeds the skin depth of this metal \cite{Cuadrado}.
Therefore, the contrast between ridges and valleys of the fingermarks is due to reflection happening at 
different path-lengths of light propagating in air. In contrast, red and orange light are
significantly transmitted through  the chalcogenide-glass CTF \cite{RJMP} as well as through
the fingermark residue (see the first and third columns in Figs. \ref{fig1}--\ref{fig3}). Therefore, the contrast between ridges and valleys of the fingermarks is also due to the 
different path-lengths of light propagating in the fingermark residue.}

We  conclude from the results presented here that the CTFs deposited on CDs and  DVDs can enhance the visual quality of fingermarks on them very significantly. Both the latent and depleted fingermarks can be developed very well for visualization, regardless of aging for a day or three days. Consistent enhancement of quality to the maximum possible, whether
the CTF is made of a glass or a metal,  indicates the versatility of the CTF technique for development of fingermarks
 on CDs and DVDs. \red{Let us also note that these conclusions have emerged from a \textit{pilot study} according to the  International Fingerprint Research Group \cite{IFRG}, since only a few samples from just one fingermark donor (with the approval of the Penn State Institution Review Board) were examined. A more systematic study
 with at least   three donors followed by statistical analyses of multiple samples from each donor still has to be performed by a legally authorized forensic crime laboratory. }

\subsection*{Disclosures}
The authors declare no conflict of interest. No data or code were used for the research reported in this article.

\subsection* {Acknowledgments}
MF thanks LUMS for one year of sabbatical leave and the Department of Engineering Science and Mechanics (Penn State) for hosting him during this period.   RAF and AL thank the Charles Godfrey Binder Endowment at Penn State for partial financial support. The  Penn State Materials Characterization Lab is acknowledged for use of the Verios G4
UC scanning electron microscope. This research was substantially funded by the US Department of Homeland Security under Grant Award No. 17STCIN00001-05-00.


\end{document}